\documentclass{article}
\usepackage[utf8]{inputenc}
\usepackage{authblk}
\usepackage{indentfirst}
\usepackage{hyperref}
\usepackage{longtable}
\usepackage{setspace}
\usepackage[margin=1in]{geometry}
\usepackage{graphicx}
\graphicspath{ {./figures/} }
\usepackage{subcaption}
\usepackage{amsmath}
\usepackage{lineno}
\usepackage{multirow}
\usepackage{color,soul}
\usepackage[version=3]{mhchem}

\usepackage[biblabel]{cite}

\title{Geometry-based Discovery of Calcium Battery Cathodes Accelerated by Foundational Machine-Learned Models}

\author[1]{Dereje Bekele Tekliye}
\author[1]{Achinthya Krishna Bheemaguli}
\author[1,*]{Gopalakrishnan Sai Gautam}

\affil[1]{Department of Materials Engineering, Indian Institute of Science, Bengaluru, 560012, India}
\affil[*]{Email: \href{mailto:saigautamg@iisc.ac.in}{saigautamg@iisc.ac.in}}

\date{}

\begin{document}

\maketitle

\begin{abstract}
\noindent Calcium batteries (CBs) are an attractive post-Li-ion technology, offering the appeal of Ca's natural abundance, redox potential close to Li, and high volumetric energy density. However, practical realization of CBs remains limited by the scarcity of positive electrode (cathode) materials that support reversible Ca$^{2+}$ (de)intercalation under typical electrochemical conditions. To address this challenge, we screen the materials project (MP) database for novel host structures that can intercalate Ca using geometry- and chemistry-based design principles. Specifically, we employ the Voronoi polyhedral volume as a descriptor of site compatibility for hosting Ca in potential frameworks. Further, we down-select candidate structures progressively through diverse criteria including charge neutrality, absence of non-Ca mobile cations, thermodynamic (meta)stability, average voltage, and Ca migration barriers ($E_m$) using foundational machine-learning (ML) models. Subsequently, we validate the ML-predicted $E_m$, obtained via an ensemble prediction among three distinct ML models, in a subset of the final candidates using density functional theory based nudged elastic band calculations. Overall, from an initial pool of 52,945 MP structures, our high-throughput workflow identifies 37 promising Ca cathode candidates, several of which exhibit favorable combinations of thermodynamic (meta)stability, voltage, and Ca$^{2+}$ mobility, marking them as strong candidates for experimental synthesis and electrochemical characterization. Particularly, we identify two Ca cathode candidates with markedly low Ca$^{2+}~E_m$, including CaSc$_2$V$_2$O$_8$ (213~meV) and CaVSO$_4$F$_3$ (376~meV), and four cathode candidates (Ca$_3$(CoO$_2$)$_4$, Ca$_3$Mn$_4$(TeO$_6$)$_2$, CaVF$_5$, and CaVSO$_4$F$_3$) with thermodynamic stability in their charged composition. Beyond identifying Ca-cathodes, our work establishes geometry-based descriptors and ML-based workflows as transferable methods for high-throughput screening, enabling the rational discovery of novel materials for battery and other applications.
\end{abstract}


\section{Introduction}
\noindent Lithium-ion batteries (LIBs) are the workhorse of modern energy storage, powering everything from portable electronics to electric vehicles and grid systems.\cite{whittingham2014ultimate,manthiram2020reflection,larcher2015towards,cano2018batteries} The high energy density, long cycle life, and rapidly declining costs of LIBs have enabled their widespread deployment at scale.\cite{tarascon2022material,nykvist2015rapidly}  Nevertheless, rising demand for energy storage, supply-chain constraints, safety considerations, and performance bottlenecks continue to motivate the search for next-generation electrode materials and chemistries.\cite{harper2019recycling, masias2021opportunities} In this context, multivalent battery systems, such as those based on Mg, Ca, Zn, or Al, offer significant promise because multi-electron redox and the possibility of using a metallic anode can facilitate higher volumetric energy density compared to LIBs.\cite{ponrouch2019multivalent, canepa2017odyssey,  blanc2020scientific, muldoon2014quest, ponrouch2016towards, arroyo2019achievements} Among multivalent systems, Ca is particularly attractive due to $i$) its redox potential being close to that of Li ($-2.87$~V vs.\ SHE), $ii$) Ca metal anode enabling high volumetric energy densities (theoretical capacity of 1337 mAh g$^{-1}$ and ~2073 Ah L$^{-1}$), and $iii$) Ca's availability on the earth's crust as the fifth most abundant element.\cite{wang2018reversible, ji2021recent, zhao2022calcium, dompablo2016joint, tchitchekova2018, ponrouch2019multivalent, arroyo2016quest, torres2019evaluation} Thus, development of reversible battery systems based on Ca as the electroactive ion retain a lot of promise as an alternative to LIBs.

Recent advances in the design of Ca electrolytes have enabled reversible Ca plating and stripping at or near room temperature, markedly improving practical feasibility of a Ca-battery (CB).\cite{shyamsunder2019reversible, wang2018plating, forero2020understanding} Yet, the positive electrode (cathode) remains the critical bottleneck for the development of a practical Ca-battery. Unlike monovalent Li$^{+}$, divalent Ca$^{2+}$ exhibits poor diffusion in most inorganic frameworks due to its large ionic radius ($\sim$1~\AA{} in an octahedral coordination environment surrounded by O$^{2-}$) and high electrostatic charge, which require `wide' diffusion pathways. Consequently, many candidate hosts exhibit high migration barriers ($E_m$),\cite{rong2015materials, hosein2021promise, chando2023exploring, jeon2022bilayered, du2025recent} structural instabilities at operating voltages, and/or undesirable reactivity with the electrolyte.\cite{gummow2018calcium} The central challenge, therefore, is to identify cathode frameworks that simultaneously deliver low Ca$^{2+}$ $E_m$, thermodynamic and interfacial stability, and practical voltages, while remaining synthesizable and chemically robust under realistic battery conditions.

Extensive experimental and computational efforts have been devoted to exploring potential Ca-battery cathodes across various crystal frameworks, including tunnel, layered, and spinel oxides,\cite{gautam2015first, jeon2022bilayered, zhang2022towards, richard2023ultra, xu2019bilayered, lu2021searching, black2022elucidation, chando2023exploring, cabello2018applicability, tojo2018electrochemical, chae2020calcium, vo2018surfactant, cabello2016advancing, park2021layered, tchitchekova2018, wang2022cav6o16} Chevrel-type sulfides,\cite{smeu2016theoretical} Prussian blue analogues,\cite{kuperman2017high, tojo2016reversible, shiga2015insertion, padigi2015potassium,  lipson2015rechargeable} polyanionic hosts such as phosphates, and sodium superionic conductor (NaSICON) structures,\cite{lipson2017calcium, kim2020high, jeon2020reversible, xu2021new, tekliye2022exploration} and fluorides.\cite{tekliye2024fluoride} For example, a high-throughput density functional theory (DFT\cite{hohenberg1964inhomogeneous, kohn1965self}) study by Lu \textit{et\,al.} identified post-spinel CaV$_2$O$_4$ and layered CaNb$_2$O$_4$ as promising compositions based on intercalation voltage, thermodynamic stability, and $E_m$ criteria, with early experiments indicating potential yet still requiring framework optimization.\cite{lu2021searching} A recent work on post-spinel CaMn$_2$O$_4$ phase showed limited cycling performance (52 mAh g$^{-1}$ at C/33),\cite{chando2023exploring} while a water-free $\beta$-Ca$_{0.14}$V$_2$O$_5$ delivered a reversible capacity of $\sim$247 mAh g$^{-1}$ with improved cycling stability and minimal dimensional change compared to hydrated and layered V$_2$O$_5$ phases.\cite{richard2023ultra}
    
Among polyanionic frameworks, FePO$_4$ accommodates $\sim$0.2 mol of Ca$^{2+}$ at 2.9 V with some reversibility.\cite{kim2020high} Fluorinated vanadium phosphate, Na$_{0.5}$VPO$_{4.8}$F$_{0.7}$, exhibits excellent Ca cycling stability, retaining 90\% of its capacity after 500 cycles with a capacity near 87 mAh g$^{-1}$.\cite{xu2021new} A recent experimental and computational study on Ca$_x$NaV$_2$(PO$_4$)$_3$ revealed a reversible Ca content limited to $x \approx 0.65$, associated with phase separation into Na-rich and Ca-rich domains.\cite{blanc2023phase} Notably, presence of Na within the framework is attributed to facilitate neighboring Ca$^{2+}$ migration, thus promoting reversible electrochemical activity, a recurring feature in studies involving polyanionic frameworks as Ca-cathodes.

Additionally, given the similar ionic radii of Ca$^{2+}$ and Na$^{+}$, structural frameworks that accommodate reversible Na intercalation are potential hosts for Ca insertion as well.\cite{shannon1969effective, shannon1976revised} Indeed, experimental studies on NaSICON-type $\mathrm{NaV_2(PO_4)_3}$ have demonstrated reversible Ca intercalation, up to 0.6 mol Ca$^{2+}$ at 3.2~V vs.~Ca,\cite{kim2020high} indicating that NaSICON frameworks can serve as hosts for Ca-cathodes, thereby motivating our prior extensive DFT-based screening of NaSICON and fluoride chemistries as Ca-cathodes.\cite{tekliye2022exploration, tekliye2024accuracy,  tekliye2024fluoride} Specifically, we evaluated the average Ca-intercalation voltage, 0~K thermodynamic stability of intercalated (or discharged) and deintercalated (charged) compositions, and Ca$^{2+}~E_m$ within the NaSICON polyanionic family, $\mathrm{Ca}_x\mathrm{M}_2(\mathrm{ZO}_4)_3$ with $Z=$ Si, P, or S and $M=$ Ti, V, Cr, Mn, Fe, Co, or Ni.\cite{tekliye2022exploration} Subsequently, we identified $\mathrm{Ca}_x\mathrm{V}_2(\mathrm{PO}_4)_3$, $\mathrm{Ca}_x\mathrm{Mn}_2(\mathrm{SO}_4)_3$, and $\mathrm{Ca}_x\mathrm{Fe}_2(\mathrm{SO}_4)_3$ as promising Ca-cathode candidates, while silicate NaSICONs were predicted to be unstable, with experimental realization of the Mn- and Fe-NaSICON candidates still pending.\cite{tekliye2022exploration} 

We extended our exploration to transition metal (TM) fluorides, including weberite-$\mathrm{Ca}_x\mathrm{M}_2\mathrm{F}_7$ and perovskite-$\mathrm{Ca}_x\mathrm{MF}_3$ compositions, using the same suite of 3$d$ TMs and computing similar metrics as the NaSICON study.\cite{tekliye2024fluoride} Importantly, we identified weberite-$\mathrm{Ca}_x\mathrm{Cr}_2\mathrm{F}_7$ and weberite-$\mathrm{Ca}_x\mathrm{Mn}_2\mathrm{F}_7$ as promising Ca-cathodes. Note that the design principle of cation substitution being largely governed by size compatibility can be generalized further by quantifying the similarity in ionic sizes via estimation of the Voronoi polyhedral volume (VPV\cite{brostow1978construction}). Unlike tabulated ionic radii, VPV captures the true local coordination environment that can accommodate a given cation, and better captures similarities across multiple combinations of structures, chemistries, and ions of interest, enabling physically-grounded identification of compatible host sites across diverse structures. We utilize the VPV metric in identifying novel Ca-cathodes in this work.

In general, DFT-based high-throughput explorations are well suited for materials discovery and design, such as the identification of potential Ca-cathodes. However, the computational costs associated with such high-throughput studies scale rapidly with increasing the chemical or structural space and the number of properties that are used during the screening process. Recent advances in machine-learned interatomic potentials (MLIPs), resulting in `foundational' or `universal' potentials, have transformed materials modeling by combining near quantum-mechanical accuracy with broad scalability and lower computational costs, enabling high-throughput and multiscale simulations.\cite{batatia2022mace, batatia2025foundation, unke2021machine, choi2025perspective, xia2025evolution, batatia2025design, neumann2024orb, rhodes2025orb} Among the available MLIPs, the multi atomic cluster expansion (MACE\cite{batatia2022mace,batatia2025foundation}) framework has emerged as a particularly promising approach as it is built on high body-order equivariant message-passing graph neural networks. Batatia et al.\cite{batatia2022mace,batatia2025design} introduced pre-trained MACE-MP-0 foundation models, namely, small, medium, and large, that differ in the degree of equivariance during message passing, i.e., $L_\text{max}$~=~0, 1, and 2, respectively. The MACE-MP-0 models are trained on the materials project (MP\cite{horton2025accelerated}) trajectory (or MPtrj\cite{deng2023chgnet}) dataset covering 89 elements, which enable accurate and transferable modeling of atomic interactions across diverse material systems. Thus, the development of MACE-MP-0, and other universal potentials, have accelerated materials discovery across a wide range of applications, including batteries,\cite{winter2023simulations, bertani2025atomic, deng2025accelerating, bheemaguli2026evaluation} catalysis,\cite{lee2025machine} photovoltaics,\cite{arber2025ion, kavanagh2025identifying} semiconductors,\cite{caro2023machine} and structural systems,\cite{lee2025accelerating} \cite{zeni2025generative} by providing a pathway to sample across more extensive chemical and structural spaces efficiently.\cite{choyal2025exploration, seth2025investigating}

Here, we integrate a geometry-based design strategy with a MLIP-accelerated high-throughput workflow to move beyond framework-by-framework evaluation and systematically identify previously unexplored chemistries and structures that can host Ca and function as cathodes for Ca-batteries. We examine available frameworks in the MP, a database of DFT–calculated properties largely based on structures from the inorganic crystal structure database,\cite{hellenbrandt2004inorganic} with a focus on identifying novel non-Ca-containing structures that can host Ca. We use the MACE-MP-0 large foundational model for rapid geometry optimization, accelerating the evaluation of the ground-state Ca-configuration and the associated total energy, and aiding the estimation of Ca $E_m$ to identify potential candidates. Note that $E_m$ predictions by a single foundational MLIP may not be as accurate across a wide range of chemistries and $E_m$ values, as suggested by a recent benchmarking study.\cite{bheemaguli2026evaluation} Thus, in addition to MACE-MP-0, we employ another foundational MLIP, namely Orb-v3,\cite{rhodes2025orb} and a graph-based transfer-learned (TL) property predictor model\cite{devi2024optimal, devi2025literature, devi2026leveraging} to increase the confidence in $E_m$ predictions and identify candidates with higher precision. The three machine-learned (ML) models operate in a `mixture-of-experts' (MoE) or an `ensemble' combination, thereby boosting the accuracy in $E_m$ predictions. 

Overall, starting from 52,945 inorganic compounds, we apply chemically-motivated filters, including the geometric Ca–VPV tolerance, charge neutrality, thermodynamic stability (at 0~K), average Ca-intercalation voltage, and Ca$^{2+}$ $E_m$, and arrive at a set of 37 promising frameworks as Ca-intercalating cathodes. Further, we validate our MoE-predicted Ca$^{2+}$ $E_m$ values with DFT–based nudged elastic band (NEB\cite{henkelman2000improved, sheppard2008optimization}) calculations, with all converged results falling within our predefined tolerance, confirming the reliability of the ensemble-guided screening. Importantly, we identify two cathode candidates that display exceptionally low DFT-NEB-calculated Ca$^{2+}~E_m$, namely CaSc$_2$V$_2$O$_8$ (213~meV) and CaVSO$_4$F$_3$ (376~meV), besides highlighting Ca$_3$(CoO$_2$)$_4$, Ca$_3$Mn$_4$(TeO$_6$)$_2$, CaVF$_5$, and CaVSO$_4$F$_3$ as promising due to their thermodynamic stability at the charged composition. Our unified high-throughput ML-accelerated discovery framework delivers orders-of-magnitude acceleration in materials discovery while maintaining the fidelity required for experimental down-selection. To our knowledge, our study is the largest-scale exploration of possible Ca-cathode chemical space, so far, and our high-throughput workflow is readily transferable for identifying novel materials for other battery chemistries and other applications.

\section{Methods}
\begin{figure}[h!]
\centering
\includegraphics[width=0.87\textwidth]{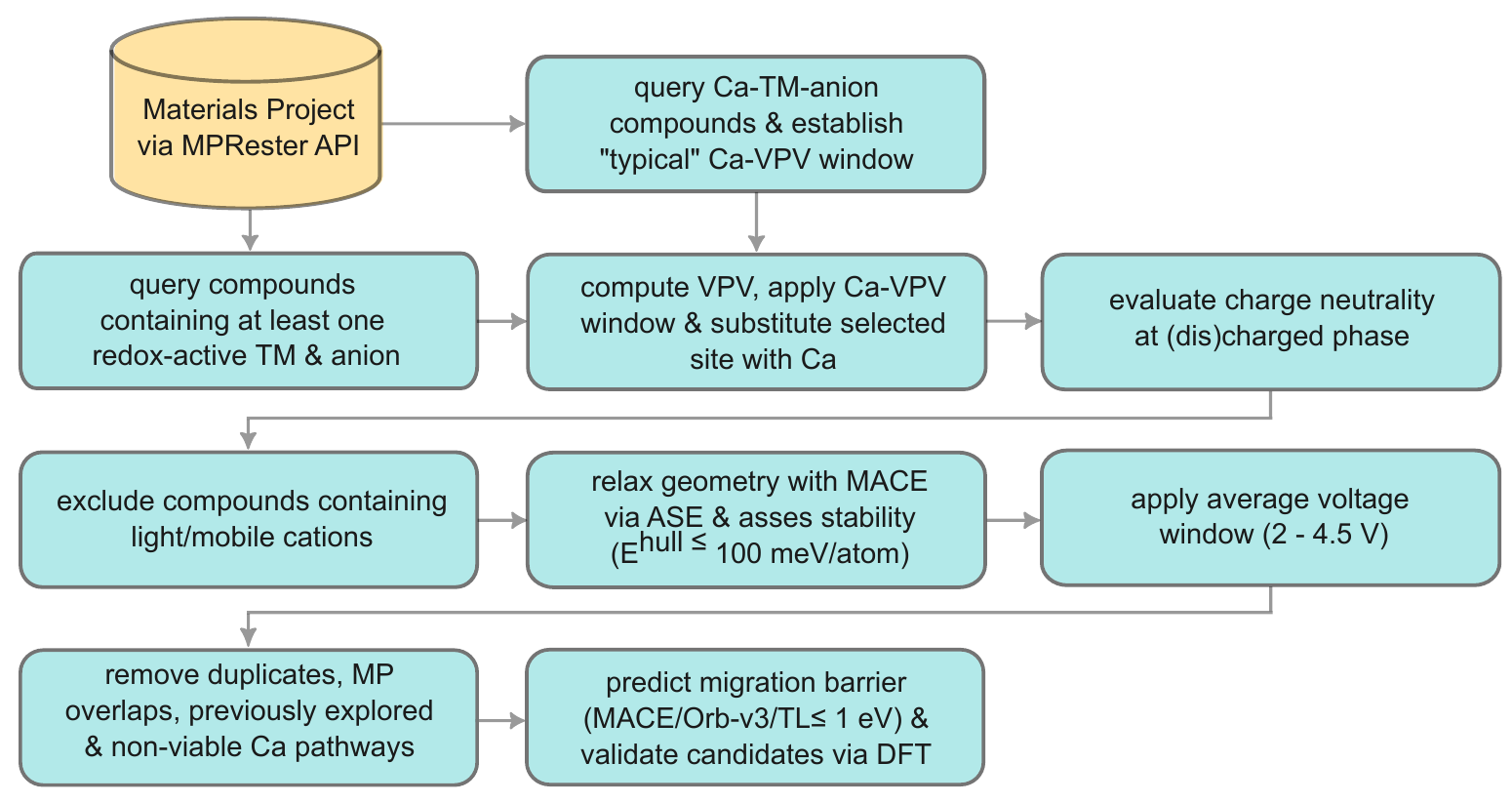}
\caption{Schematic overview of the high-throughput ML-accelerated screening workflow developed in this work for identifying promising Ca-intercalation cathode frameworks.}
\label{fig:workflow}
\end{figure}

\subsection{VPV-based framework screening}
\noindent \textbf{Figure~\ref{fig:workflow}} summarizes the screening workflow used in this study. We began by querying Ca-containing compounds that include at least one TM (not necessarily redox-active) and one anion from the MP database (as available on May 7, 2025), using the \texttt{MPRester} API in \texttt{pymatgen}.\cite{ong2013python} We refer to the collection of all queried compounds, amounting to 4,350 frameworks, as `set 1' and considered compounds regardless of whether they have been previously studied for battery applications. Subsequently, we used the retrieved structures to estimate the `typical' polyhedral volumes that Ca$^{2+}$ occupies in Ca-containing structures, based on the VPV metric. We computed the Ca VPV in a given structure with \texttt{pymatgen}’s \texttt{VoronoiNN} class by constructing the Voronoi tessellation around all Ca sites and averaging the corresponding polyhedral volumes. The VPV calculation is done to establish a representative Ca–VPV window, i.e., the typical range of polyhedral volumes that Ca prefers among inorganic structures, centered around the median VPV value with a $\pm$5\% tolerance across all queried compounds that contain Ca. This VPV range was then used to quantify geometric compatibility of lattice (or void) sites that can possibly accommodate Ca and hence identify potential Ca sites within structures that do not contain Ca \textit{a priori}.   
    
Upon obtaining the typical Ca VPV, we queried the MP database by explicitly excluding Ca-containing entries and including at least one redox-active TM (Ti, V, Cr, Mn, Fe, Co, Ni, Nb, and/or Mo), and at least one anion (O, S, Se, Te, N, P, F, Cl, Br, and/or I) to identify novel cathodes that do not contain Ca \textit{a priori}, forming our `set 2' containing 52,945 compounds. The list of redox-active TMs, the corresponding allowed oxidation states of TMs, and anions considered is compiled in \textbf{Table~S1} of the supporting information (SI). Any cation species other than the `framework-forming' elements (i.e., the redox-active TM and anions) were considered for potential substitution with Ca, subject to their compatibility with the VPV metric, along with possible interstitial/vacancy sites in the structure. If no species other than the redox-active TM and anions were present in a structure, we assessed possible interstitial/vacancy sites available in the structure for Ca insertion. To identify potential vacancy/interstitial sites in a given structure, we used \texttt{pymatgen}’s \texttt{TopographyAnalyzer}. We required any candidate vacancy/interstitial site to have a minimum distance of 1~\AA{} to the nearest framework atom. If multiple sites that are closer than 1~\AA{} from one another are identified, we retained only a single representative site among them. Upon identification of a suitable site, we placed a dummy chemical species (`X') and calculated the VPV for that site.
    
Finally, we used the established Ca–VPV tolerance window to estimate the geometric compatibility of possible substitution and/or vacancy/interstitial sites across all non-Ca-containing queried compounds of set 2. If a structure contained both substitutional and X sites that can be occupied by Ca, we considered the possibility of Ca occupying the substitutional site only. Thus, X sites are considered in the absence of any possible substitutional sites that Ca can occupy in a framework. If multiple sites in a structure fell within the VPV tolerance window, we selected the site with a VPV value closest to the Ca-VPV median for Ca substitution/insertion. In the event of a tie, we chose a site at random. Note that we did not down-select any structures from set 1 that contain redox-active TMs and are within the target Ca-VPV window since such structures may have been considered already in previous screening studies.\cite{lu2021searching} Thus, our final list of geometry-compatible compounds contain frameworks from set 2 that fall within the Ca-VPV window, with possible Ca substitution at existing cation sites or Ca insertion in X sites. The queried list of compounds (i.e., MP-IDs of structures) comprising our sets 1 and 2 are available in our \href{https://github.com/sai-mat-group/high-throughput-ca-cathodes}{GitHub} repository.

\subsection{Charge neutrality constraint}
\label{sec:methods-charge}
\noindent Charge neutrality is one of the prerequisites for thermodynamic stability and, by extension, for chemical synthesizability. As cathode compositions evolve during electrochemical cycling, we require that each composition encountered remain electrostatically neutral. We therefore evaluate the two end members, namely, the fully `discharged' state containing the maximum possible occupation of Ca and the fully `charged' state with the minimum possible occupation of Ca. For both end member compositions, we constrain the sum of ionic charges, as determined via allowed oxidation states for different species present (see \textbf{Table~S1}), to be zero. We assigned oxidation states using \texttt{pymatgen}’s \texttt{Composition} class, following chemical constraints on possible oxidation states (see Section~S1 of SI).\cite{greenwood2012chemistry} For instance, we limit the oxidation states of redox-active TMs to be in the range of +2 to +4 (except V which is allowed between +2 and +5) and allowed fractional TM oxidation states (e.g., 3.5+) to account for mixed-valence. On the other hand, we allowed either a single or a small set of integer oxidation states for \textit{p}-block elements. Among the queried structures, we did not encounter cases where both positive and negative oxidation states had to be assigned for the same element (e.g., both +6 and -2 for S in the same structure). When multiple \textit{p}-block elements are present, we constrained the oxidation state of the most electronegative element to be in its lowest negative state (e.g., -2 for O). While our oxidation state constraints are not exhaustive, they do eliminate most charge-imbalanced compositions, with any remaining cases removed by subsequent thermodynamic evaluation (see Section~\ref{sec:methods-stability}).

\subsection{Mobile-cation–containing compound exclusion}
\label{sec:methods-mobile}
\noindent Several cations, including H, Li, Na, K, Mg, and Zn, are known to be mobile in solid-state materials, particularly cathode compositions.\cite{canepa2017odyssey, shinde2023li, xu2022electrochemical, wu2024challenges} In structures that contain both Ca and an additional mobile cation, such as Li, the electrochemical behavior can arise from the motion of Ca and/or Li. Therefore, to ensure that Ca serves as the sole electroactive ion in a candidate cathode material, all Ca-cathode candidates that satisfy the charge-neutrality constraint but contain additional mobile cations (as mentioned above) were excluded from the dataset. The remaining charge-neutral candidates that contain only Ca as the electroactive species are evaluated subsequently.

\subsection{Thermodynamic stability and average voltage}
\label{sec:methods-stability}
\noindent To evaluate thermodynamic stability with respect to other competing phases, we use the MACE-MP-0 large foundational model, hereafter referred to simply as MACE. Specifically, we performed MACE-based geometry relaxations for the selected candidates following Sections~\ref{sec:methods-charge} and \ref{sec:methods-mobile}, in both their charged and discharged compositions. We employed the atomistic simulation environment (ASE\cite{hjorth2017atomic}) for structural optimization, using the MACE calculator for estimating total energies and atomic forces. ASE’s \texttt{UnitCellFilter} was used to allow simultaneous relaxation of both atomic positions and cell vectors, via  the \texttt{QuasiNewton} optimizer. We considered a structural optimization converged when the interatomic forces were below $|0.05|$~eV/Å over a maximum of 1,000 optimization steps. 

Because the MACE model used in this work is trained on uncorrected MP energies, we applied the \texttt{MaterialsProject2020Compatibility}\cite{jain2011formation, wang2021framework} correction scheme in \texttt{pymatgen} to our MACE-calculated energies to maintain consistency with the data available on MP. Specifically, we used the corrected MP energies of all possible competing phases, as available on MP, for constructing the corresponding 0~K convex hulls and assessing the thermodynamic stability for both charged and discharged compositions. Notably, we computed the energy above the convex hull (E$^\mathrm{hull}$) for each Ca-based compound. Given typical DFT errors in 0~K stability predictions\cite{sun2016thermodynamic} and residual MACE-to-DFT deviations, we used a synthesizability threshold of E$^\mathrm{hull}$ $\leq$ 100~meV/atom for the new charged/discharged compositions considered, i.e., compounds with E$^\mathrm{hull}$ $\leq$ 100~meV/atom were considered  (meta)stable. 

For each (meta)stable compound, we subsequently computed the average intercalation voltage from the total energy difference between the discharged and charged compositions using the approximate Nernst equation, as shown in Equation~(\ref{eq:voltage})\cite{aydinol1997ab},
\begin{equation}
V = -\frac{\Delta G}{2Fx} \approx -\frac{E_\mathrm{discharged} - E_\mathrm{charged} - x\,\mu_\mathrm{Ca}}{2Fx}
\label{eq:voltage}
\end{equation}
where $E_\mathrm{discharged}$ and $E_\mathrm{charged}$ are the corrected MACE-calculated total energies of the discharged and charged structures, respectively. $x$ is the number of intercalated Ca atoms per formula unit of a given charged composition, $F$ is the Faraday constant, and $\mu_\mathrm{Ca}$ is the chemical potential (energy per atom) of metallic calcium in its ground-state face-centered-cubic phase. The Gibbs energy change in Equation~\ref{eq:voltage} is approximated by the total energy difference ($\Delta G \approx \Delta E$), neglecting $pV$ and entropic contributions, consistent with typical high-throughput DFT workflows.

\subsection{Structure curation for Ca-mobility estimation} 
\noindent In order to uniquely identify candidate structures for the computationally-intensive Ca-mobility estimations, we removed duplicacy among the screened frameworks by first grouping entries based on composition in the discharged state. Subsequently, we clustered the corresponding structures within each group based on crystallographic similarity using \texttt{pymatgen}'s \texttt{StructureMatcher} with its default tolerances (i.e., being symmetry- and scale-aware, species-sensitive). From each cluster of `similar' structures, we retained the entry with the lowest $E^\mathrm{hull}$, yielding a non-redundant set of compositionally unique structures with high thermodynamic (meta)stability.

As an additional redundancy check, we cross-examined the structural similarity of candidate structures obtained against available MP structures using \texttt{StructureMatcher} (default tolerances). We grouped our screened structures by composition in the discharged state and subsequently compared the discharged structures within each each group with available MP structures with the same composition. We eliminated identified candidates that matched with structures in the MP and were previously studied as possible Ca cathodes (e.g. CaV$_2$(PO$_4$)$_3$,\cite{tekliye2022exploration, kim2020high} CaV$_2$O$_5$,\cite{gautam2015first}  CaFe$_2$(SO$_4$)$_3$\cite{tekliye2022exploration}). On the other hand, MP-matched entries without prior Ca-cathode studies were retained for Ca-mobility evaluation. Note that we also removed compounds lacking viable Ca (de)insertion pathways for subsequent Ca-mobility calculations. Operationally, we  defined structures lacking viable Ca (de)insertion pathways as those that exhibit ($i$) an absence of any pair of Ca-sites with hopping distances of $< 6$~\AA,\ or ($ii$) the absence of a percolating network of Ca-sites enabling insertion/extraction as determined via visual inspection.

\subsection{Migration barrier calculations}
\noindent For selected frameworks with `small' unit cells (i.e., lattice parameters $< 8$~\AA), we constructed larger supercells of lattice parameters at least 8~\AA\ to avoid spurious interactions between the migrating Ca$^{2+}$ and its periodic images. To assess Ca$^{2+}$ mobility, we analyzed each discharged structure (unit or supercell) by explicitly visualizing and enumerating the relevant vacancy-mediated migration pathways, typically amounting from one to four per structure. Additionally, we recorded whether long-range transport of Ca required all of these pathways or only a subset to be active. To identify Ca-cathode candidates with acceptable rate performance, we consider a fairly liberal maximum allowed $E_m \approx 1$~eV, which corresponds to a possibly reversible Ca (de)intercalation across cathode particles of $\sim$5~nm characteristic size, and operating at 333~K (60~°C) and at C/10 rate.\cite{lu2021searching, tekliye2022exploration, tekliye2024fluoride} Thus, we considered Ca$^{2+}$ migration pathways exhibiting $E_m \leq 1$~eV to be electrochemically active.

For each pathway within a given structure, endpoint configurations (i.e., with the Ca$^{2+}$ located at the start or the end sites) were generated and geometry-optimized using MACE with ASE's \texttt{QuasiNewton} algorithm upto a force threshold of $|0.05|~ \mathrm{eV\,\AA^{-1}}$. Corresponding minimum-energy paths were computed using the NEB method in conjunction with MACE and the Broyden--Fletcher--Goldfarb--Shanno (BFGS\cite{broyden1970convergence}) algorithm with a spring constant of \(2.5~\mathrm{eV\,\AA^{-2}}\) between images along the band. For each structure, we generated an initial band of nine images (seven intermediate images plus the endpoints) using the image-dependent pair potential\cite{smidstrup2014improved} interpolation scheme, which was found to provide a better initial guess for migration pathways compared to linear interpolation in our previous benchmarks.\cite{bheemaguli2026evaluation} We considered the NEB calculations converged if the perpendicular component of the band force fell below \(\left|0.05\right|~\mathrm{eV\,\AA^{-1}}\) or the NEB optimization reached a maximum of 1{,}000 steps. The $E_m$ was taken as the energy of the highest (saddle-point) image relative to the lowest energy image.

Alongside MACE, we used the Orb-v3 foundational model (conservative variant) as an independent comparator to cross-validate predictions and guide candidate down-selection for DFT-based NEB validation. The conservative Orb-v3 variant computes forces and stresses as analytic derivatives of the energy via backpropagation (which is desirable for NEB), and is designed for fast and accurate prediction across diverse inorganic materials. Notably, we considered MACE and Orb-v3 foundational models in this work based on prior benchmarking that demonstrated strong agreement of these models with DFT-computed $E_m$.\cite{bheemaguli2026evaluation} For Orb-v3 based calculations, we  used the corresponding ASE calculator to evaluate $E_m$ with the same NEB settings as for MACE, except that the endpoint structures were relaxed with the \texttt{BFGS} optimizer (rather than \texttt{QuasiNewton}) due to faster convergence based on our preliminary tests. 

In addition to foundational MLIPs, we used a transfer-learning based property predictor graph model built on the ALIGNN architecture (hereafter referred to as TL) to predict $E_m$.\cite{devi2026leveraging, devi2024optimal} ALIGNN encodes atomic, bond, and bond-angle information via edge-gated graph convolutions to produce a pooled crystal embedding that is mapped to $E_m$ by a multilayer perceptron.\cite{choudhary2021atomistic} The TL model employed here was previously fine-tuned on a literature-derived dataset of 619 DFT-calculated $E_m$.\cite{devi2025literature} For each migration pathway considered, we provided a pathway-encoded input in which the migrating ion occupies the sites corresponding to the initial and final configurations together with three linearly interpolated intermediate images (i.e., an NEB-style initialization). With this band-like structural input, the TL model predicts the corresponding $E_m$ for that pathway, which is used in conjunction with MACE and Orb-v3 predicted $E_m$ values for DFT-NEB calculations.

\subsection{DFT-NEB validation}
\noindent For the candidates identified as exhibiting promising Ca mobility, based on MACE, Orb-v3, and TL model estimations, we validated the predictions using DFT-based NEB calculations.\cite{henkelman2000improved, sheppard2008optimization} To generate the initial band for a DFT-NEB calculation, we first performed a short MACE–NEB run with five intermediate images to generate an initial Ca-migration pathway, leveraging the ability of MACE to provide an accurate geometric description of the band.\cite{bheemaguli2026evaluation} The corresponding endpoints (i.e., the initial and final images) were subsequently relaxed at the DFT level, and combined with the MACE-optimized intermediate images as the initial band for the DFT-NEB calculation. Thus, the (super)cell sizes used for DFT-NEB calculations were identical to prior (super)cells used for MACE, Orb-v3, and TL calculations.

All spin-polarized endpoint relaxations were performed with DFT using the Vienna \textit{ab initio} simulation package\cite{kresse1993ab, kresse1996efficient} and the Perdew--Burke--Ernzerhof functionalization of the generalized-gradient approximation\cite{perdew1996generalized} for electronic exchange and correlation, to minimize computational costs and mitigate convergence difficulties.\cite{devi2022effect} We employed projector augmented-wave\cite{kresse1999ultrasoft, blochl1994improved} potentials with frozen cores, used a plane-wave kinetic-energy cutoff of 520~eV, and sampled the irreducible Brillouin zone using $\Gamma$-centered Monkhorst--Pack\cite{monkhorst1976special} meshes with a minimum $k$-point grid density of 16 $k$-points per \AA. Endpoint relaxations were carried out at fixed cell shape and volume, without symmetry constraints, and with convergence thresholds of $10^{-5}$~eV for the total energy and $|0.03|$~eV\,\AA$^{-1}$ for the atomic forces. The DFT-powered NEB calculations used a spring constant of 5~eV\,\AA$^{-2}$ between the adjacent images, with the NEB calculation considered converged when the maximum perpendicular force on the band fell below $|0.05|$~eV\,\AA$^{-1}$.

\section{Results}
\subsection{Querying MP and target Ca–VPV window}
\begin{figure}[h!]
    \centering
    \includegraphics[width=1\textwidth]{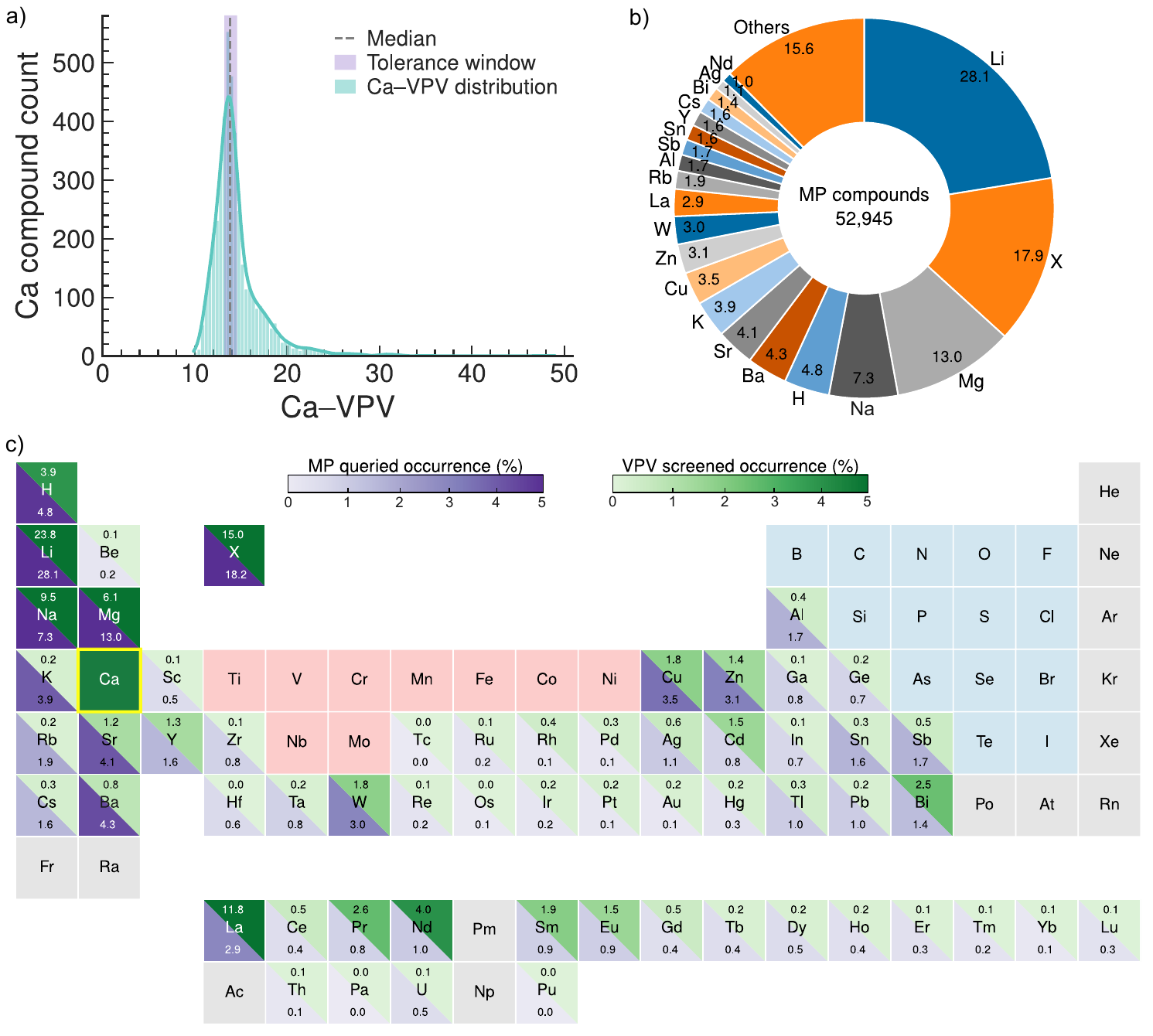}
    \caption{a) Distribution of Ca-VPV in Ca-containing compounds, as queried from MP. b) Distribution of queried MP compounds that do not contain Ca, grouped by the presence of potential Ca-substitution sites. c) Section of the periodic table showing the distribution of potential host frameworks that can accommodate Ca. The green heat map (upper-right triangles) indicates the total number of substitutional sites originally occupied by a given element, after matching with the Ca-VPV criterion. The purple heat map (lower-left triangles) shows the number of corresponding element-containing compounds in the  full queried dataset, as represented in panel b. X represents vacant/interstitial sites that can be replaced with Ca.}    
    \label{fig:mp-distribution}
\end{figure}

\noindent \textbf{Figure~{\ref{fig:mp-distribution}}a} represents the distribution of local environments containing Ca, as characterized by the Ca-VPV metric, among the 4,350 structures that comprise set 1 of MP-queried frameworks. Notably, we observe a tight clustering of Ca-containing local environments (blue shaded region) around the median Ca-VPV value of 13.86~\AA$^3$ (vertical dashed line in \textbf{Figure~{\ref{fig:mp-distribution}}a}). The minimum and maximum value of Ca-VPV within our queried set are 9.85 and 49.00~\AA$^3$, respectively. Additionally, our tolerance window of $\pm$5\% around the median gives us a target Ca-VPV window of 13.17-14.56~\AA$^3$ (purple band in \textbf{Figure~{\ref{fig:mp-distribution}}}a) for determining geometric compatibility of possible sites to host Ca in novel frameworks (i.e., within set 2). Note that we use the $\pm$5\% tolerance window to capture natural variations in Ca coordination while keeping a tight geometric criterion to determine compatibility with a `typical' coordination environment preferred by Ca. 

To provide a quantitative overview of the availability of host frameworks that can accommodate Ca, we present the percentage distribution of queried candidate structures (totally 52,945) that form set 2 in \textbf{Figure~\ref{fig:mp-distribution}b}. The structures are grouped by potential Ca-substitution sites, where X denotes vacancies/interstitials. As expected, Li-containing compounds dominate the distribution, accounting for ~28.1\%, reflecting the statistically large number of Li compounds present in MP. Li is followed by compounds with X sites (17.9\%), Mg (13.0\%), Na (7.3\%), H (4.8\%), Ba (4.3\%), Sr (4.1\%), and K (3.9\%). There are several compounds containing cations such as Cu, Zn, W, etc., as possible substitutional sites for Ca, accounting for $\leq$3.5\% of the distribution. Compounds containing potential substitutional cations that account for $<$1\% of the queried data are grouped and represented as ``Others" (accounting for a total of 15.6\% of the dataset). As a start of our screening approach (\textbf{Figure~{\ref{fig:workflow}}}), all structures contributing to the distribution in \textbf{Figure~{\ref{fig:mp-distribution}}b} are passed through the Ca-VPV compatibility filter.

\subsection{Geometric criterion screening}
\noindent Upon applying the Ca-VPV tolerance window (\textbf{Figure~{\ref{fig:mp-distribution}}a}) as a geometric compatibility check, our full dataset reduces from a total of 52,945 initial structures to 5,946 candidate compounds, highlighting the limited structural compatibility within the inorganic chemical space to host Ca. \textbf{Figure~\ref{fig:mp-distribution}c} summarizes the statistical distribution of the VPV-matched set, shown alongside the entire queried dataset on a section of the periodic table. Cells corresponding to redox-active TMs and anions are shown as light red and light blue, respectively, and display only the element symbol (i.e., no numerical annotations), since these sites are not considered for potential replacement with Ca. Elements not present in the full dataset are either omitted or shown in gray for visual completeness. Ca, representing the substituting species, is indicated by the yellow-bordered cell shaded in dark green.

For each potential substitutional element (or X), the green heat map (upper triangles in \textbf{Figure~\ref{fig:mp-distribution}c}) encodes the percentage of the filtered compounds that contain the element, i.e., the percentage of a given element-containing structures that satisfy the target Ca-VPV window. The corresponding percentage value is printed on the top of the cell alongside the element symbol in the upper triangle. The percentage of compounds containing a given substitutional element within the entire queried dataset (i.e., prior to Ca-VPV matching) is reported in the lower triangle and highlighted by the purple heat map. The compound count and fraction of selected compounds for a given element is compiled in \textbf{Figure~S1} of the SI. 

Our data shows that Li-containing compounds dominate both the VPV-matched set (23.8\%) as well as the overall set (28.1\%), which can be attributed to the bias within MP itself. Hosts with X sites are the next most common, accounting for 15\% and 18.2\% of the VPV-matched and overall sets, respectively. Interestingly, La-containing compounds appear in only 2.8\% of the overall dataset but constitute 11.8\% of the VPV-matched set, indicating a strong enrichment among the filtered compounds and a strong similarity in local environments inhabited by La with Ca. Other notable substitutional cations include Na (9.5\% VPV-matched, 7.3\% overall), Mg (6.1\% VPV-matched, 13.0\% overall), Nd (4.0\% VPV-matched, 1.0\% overall), and H (3.9\% VPV-matched, 4.8\% overall).

\subsection{Filtering on charge neutrality}
\begin{figure}[h!]
    \centering
    \includegraphics[width=1\textwidth]{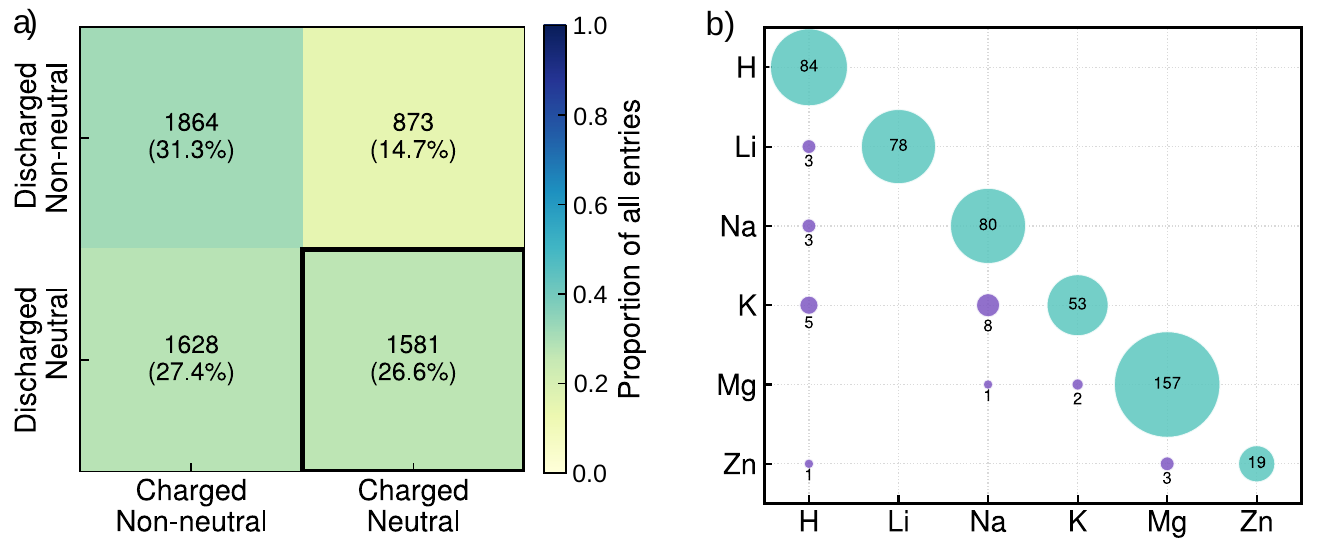}
    \caption{a) Contingency matrix of charge neutrality among discharged and charged compositions of VPV-matched compounds. b) Distribution of compounds removed due to the presence of non-Ca mobile cation(s).}
    \label{fig:charge_remove}
\end{figure}

\noindent As a pre-requisite for thermodynamic (meta)stability, we evaluate the charge neutrality of the two end-member compositions of each VPV-matched candidate framework, namely, the fully discharged and the fully charged compositions. \textbf{Figure~\ref{fig:charge_remove}a} shows a heat map of the contingency matrix for charge neutrality in the charged and discharged compositions. Among 5,946 VPV-matched compounds, 31.3\% (or 1,864) are not charge neutral in either state. Further, 27.4\% (1,628) of frameworks are charge neutral only in the discharged state and 14.7\% (873) are charge neutral only in the charged state. Finally, 26.8\% (1,581) of the VPV-matched compounds are charge neutral in both charged and discharged compositions, thus representing the strongest set of candidates for further screening. 

\subsection{Exclusion of mobile-cation–containing compounds}
\noindent To ensure that Ca remains the electroactive ion, we exclude 445 compounds (out of 1,581 charge-neutral compounds) that host additional mobile cation(s), namely, H\textsuperscript{+}, Li\textsuperscript{+}, Na\textsuperscript{+}, K\textsuperscript{+}, Mg\textsuperscript{2+}, and Zn\textsuperscript{2+}. The statistics of the compounds removed due to the presence of a non-Ca mobile cation is presented as bubble plots in \textbf{Figure~{\ref{fig:charge_remove}}b}. The size of each bubble in \textbf{Figure~{\ref{fig:charge_remove}}b} is indicative of the number of excluded structures due to the presence of other mobile cation(s). Diagonal teal bubbles indicate the count of compounds that contain only the indicated mobile cation (besides Ca). For example, 157 structures are excluded due to the presence of only Mg as an additional mobile cation. Off-diagonal purple bubbles indicate the number of compounds in which the corresponding pair of mobile cations co-exist. For instance, 8 structures are excluded due to the presence of both Na$^+$ and K$^+$ in the structure. Finally, the application of this filter results in a total of 1,136 candidate compounds that only contain Ca as an electroactive ion.

\subsection{Thermodynamic stability}
\begin{figure}[h!]
    \centering
    \includegraphics[width=1\textwidth]{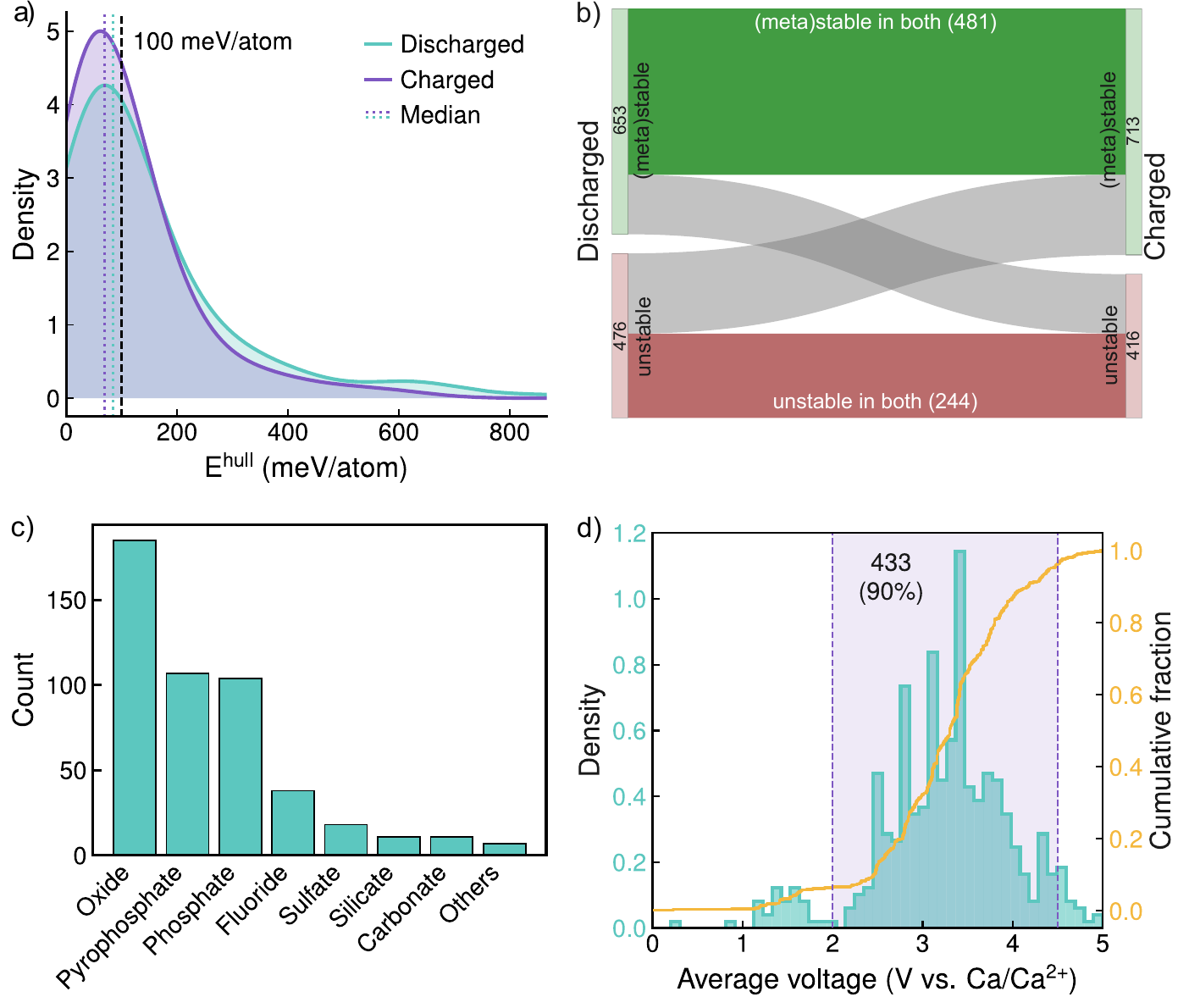}
    \caption{a) The $E^\mathrm{hull}$ distribution of the discharged (teal) and charged (purple) candidate compositions and their corresponding median values (vertical dotted lines). The black vertical dashed line marks the (meta)stability criterion ($E^\mathrm{hull}$ = 100 meV/atom). b) Sankey diagram tracing the thermodynamic (meta)stability of discharged and charged compositions. c) Screened (meta)stable frameworks grouped by chemistry. d) Computed average voltage (versus Ca metal) between the fully discharged and charged compositions for screened (meta)stable frameworks.}
    \label{fig:therm_stability} 
\end{figure}

\noindent We use $E^\mathrm{hull}$ to assess thermodynamic (meta)stability of the 1,129 screened cathode materials,\cite{sai2020exploring} with \textbf{Figure~\ref{fig:therm_stability}a} showing the distribution of MACE-computed $E^\mathrm{hull}$ values. The reduction from 1,136 compounds (after excluding structures having non-Ca mobile cations) to 1,129 compounds is due to the lack of convergence in our MACE calculations for seven structures, all of which share the common feature that Ca substitutes H in the host structure. Specifically, \textbf{Figure~\ref{fig:therm_stability}a} displays the probability density from a kernel density estimate (KDE) with units of 1/eV, as a function of $E^\mathrm{hull}$, for both discharged (teal curve) and charged (purple curve) compositions. Both discharged and charged density curves are normalized independently such that the area under each curve equals 1, with their corresponding medians indicated as teal and purple vertical dotted lines, respectively. The density at a given $E^\mathrm{hull}$ in \textbf{Figure~\ref{fig:therm_stability}a} reflects the rate of occurrence of structures around the $E^\mathrm{hull}$ value, i.e., density at $E^\mathrm{hull}$ $\times$ window of $E^\mathrm{hull}$ = fraction of compounds within that window around the $E^\mathrm{hull}$. For example, a density of 5 at $E^\mathrm{hull}$=~0.05~eV/atom on the discharged curve in \textbf{Figure~\ref{fig:therm_stability}a} indicates that 5\% of the overall 1,129 discharged structures ($\sim$56 structures) occur within a 0.01~eV/atom window around $E^\mathrm{hull}$=~0.05~eV/atom (i.e., 5$\times$0.01 = 0.05 or 5\%). The black vertical dashed line indicates the synthesizability heuristic, i.e., $E^\mathrm{hull} =$100~meV/atom,\cite{sun2016thermodynamic} used in our study to identify frameworks that are possibly synthesizable. Thus, phases on the hull ($E^\mathrm{hull}=0$) are predicted ground states, whereas those with $0 < E^\mathrm{hull} \leq 100$~meV/atom are considered metastable with respect to decomposition into other lower-energy phases. The $E^\mathrm{hull}$ range in \textbf{Figure~\ref{fig:therm_stability}a} is capped at the 99$^\mathrm{th}$ percentile of values to avoid long tails dominating the plot.

Notably, the discharged and charged compositions exhibit distinct distributions, as highlighted by their corresponding median $E^\mathrm{hull}$ values of 85 and 69~meV/atom, respectively. Additionally, the value of $E^\mathrm{hull}$ with the highest density for discharged phases is 69~meV/atom, which is slightly higher compared to the corresponding peak for the charged phases (62~meV/atom). However, a similar fraction of charged compositions ($\sim$3\%) and discharged compositions ($\sim$2\%) are predicted to be stable, i.e., $E^\mathrm{hull}=0$. Although the marginally higher occurrence of (meta)stable charged compositions compared to discharged compositions reflects our workflow of targeting non-Ca-containing structures (\textbf{Figure~{\ref{fig:workflow}}}), we do expect experimental strategies targeting the synthesis of charged compositions (with subsequent electrochemical intercalation of Ca) to be statistically more successful compared to synthesizing discharged compositions with subsequent Ca removal. Finally, we observe a non-negligible subset of candidates exhibiting $E^\mathrm{hull}$ well beyond $100$~meV/atom, indicating substantial thermodynamic driving forces for decomposition, making these phases unlikely to be accessible under conventional synthesis conditions.

While the majority of screened candidates cluster at low $E^\mathrm{hull}$ values at either their discharged or charged compositions, cathode candidates that exhibit $E^\mathrm{hull} \leq$100~meV/atom in both their discharged and charged states are promising as Ca-cathodes. To identify such candidates, we assign each screened framework to one of four categories, namely, (meta)stable in both discharged and charged states, (meta)stable only at discharged, (meta)stable only at charged, or unstable in both states, and visualize the statistics as a `Sankey' diagram in \textbf{Figure~{\ref{fig:therm_stability}}b}. The left and right nodes of \textbf{Figure~\ref{fig:therm_stability}b} sum up to the (meta)stable and unstable discharged and charged compositions, respectively, with the thickness of the colored links indicating the number of compounds that overlap in the corresponding stability class. 

Of the 1,129 frameworks evaluated, 653 discharged and 713 charged compositions are (meta)stable, as represented by light green nodes in \textbf{Figure~\ref{fig:therm_stability}b}, while 476 discharged and 416 charged compositions are unstable, as shown by light red nodes. Importantly, our stability screening identifies 481 frameworks that are (meta)stable in both their charged and discharged states, as highlighted by the green link, indicating the most promising set of Ca-cathode candidates based on our threshold $E^\mathrm{hull}$. Thus, we limit our voltage and Ca-mobility evaluations to these 481 frameworks that are (meta)stable in both their discharged and charged states. Also, we find 244 frameworks that are unstable in both discharged and charged states (red link), 172 frameworks that are (meta)stable in discharged but unstable in charged (thin grey link), and 232 frameworks that are (meta)stable in charged but unstable in discharged (thick grey link).

\textbf{Figure~\ref{fig:therm_stability}c} shows a histogram of the screened (meta)stable frameworks (i.e., 481 candidates) grouped by chemistry, namely, oxide, pyrophosphate, phosphate, fluoride, sulfate, silicate, carbonate, and others. Oxides dominate the set, followed, in decreasing abundance by pyrophosphates and phosphates. The dominance of oxides is in line with the overall data distribution in MP. Interestingly, we observe only a single chalcogenide (i.e., sulfide or selenide) host, namely CaTi(TlS)$_4$, to be selected by our screening process thus far, possibly indicating negligible occurrences of (meta)stable combinations of Ca, redox-active TM, and S/Se as the anion.

\subsection{Average voltage}
\noindent \textbf{Figure~\ref{fig:therm_stability}d} shows the distribution of average Ca intercalation voltages, versus Ca metal, calculated across the fully discharged and charged compositions, for the 481 frameworks that pass the thermodynamic (meta)stability filter. The teal bars (left y-axis) indicate the probability density of structures  at a given average voltage (computed using a KDE with units of 1/V, as obtained by normalizing the histogram such that the total area is 1) and the orange curve (right y-axis) signifies the cumulative fraction of structures over the spanned voltages. The density metric represents a rate of occurrence of structures within a voltage window, as a function of voltage, similar to the notation used in \textbf{Figure~{\ref{fig:therm_stability}}a}. We adopt a practical operating voltage window of 2.0–4.5~V (shaded region in \textbf{Figure~\ref{fig:therm_stability}d}), consistent with the electrochemical stability limits of current Ca electrolytes,\cite{ponrouch2016towards, gummow2018calcium} and identify frameworks within this range as potential Ca-cathode candidates for subsequent mobility evaluation. Similar voltage windows have been used in past cathode screening studies as well.\cite{tekliye2022exploration,tekliye2024fluoride,lu2021searching} 

Notably, 433 out of 481 frameworks (90\%) fall within the 2.0-4.5~V voltage interval, indicating substantial overlap between thermodynamic (meta)stability and electrochemical compatibility. The cumulative curve rises steeply across the 2.0–4.5~V region, reflecting this concentration of candidates in the practically relevant voltage range, with the curve flattening at both the low and high voltage ranges. The low-voltage tail below 2~V corresponds to frameworks that are more suitable as potential anodes, whereas the high-voltage tail above 4.5~V likely exceeds the oxidative stability limits of current electrolytes. Accordingly, the 433 frameworks within the 2.0–4.5~V window, which also satisfy the $E^\mathrm{hull}$ criterion, are advanced to the Ca-mobility screening (see below).

\subsection{Structure curation for mobility analysis}
\begin{figure}[h!]
    \centering
    \includegraphics[width=0.8\textwidth]{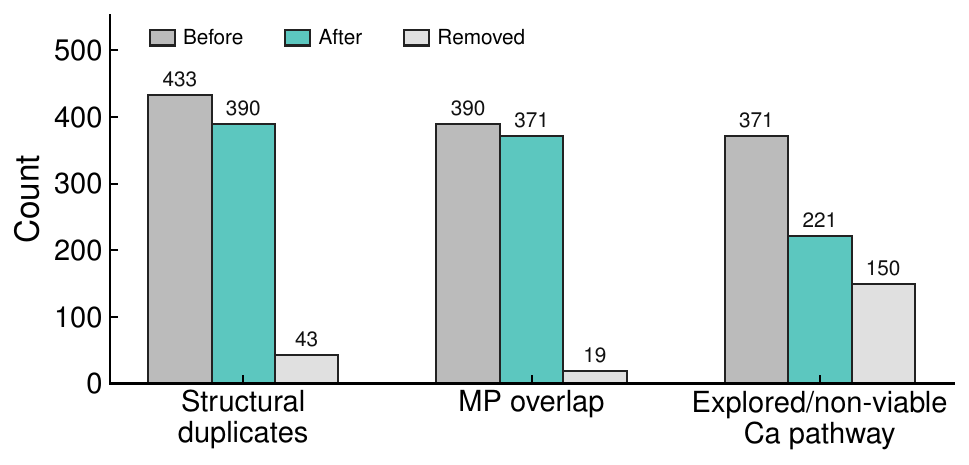}
    \caption{Sequence of data curation prior to Ca-mobility evaluation. Dark gray, teal, and light gray bars show the number of frameworks before each step, retained after the step, and removed by the step, respectively, for the three sequential filters considered.}
    \label{fig:deduplicate}
\end{figure}

\noindent \textbf{Figure~\ref{fig:deduplicate}} summarizes the progressive de-duplication and curation of candidate frameworks for Ca$^{2+}$ mobility analysis. The dark gray, teal, and light gray bars report, respectively, the number of frameworks prior to each curation step, those retained after the step, and those removed by that step, across three filters: structural de-duplication, overlap with MP entries, and the removal of frameworks corresponding to previously explored or non-viable Ca pathways. Starting from the 433 voltage-screened candidates, structural de-duplication eliminates 43 redundant entries, reducing the set to 390 unique frameworks. Subsequent removal of 19 structures that overlap with existing MP entries narrows the pool to 371 frameworks. A final filter that excludes frameworks previously explored as Ca cathodes and those lacking viable Ca (de)insertion pathways reduces the set to 221 candidates. Thus, the initial pool of 433 (meta)stable, voltage-compatible frameworks is distilled to 221 structurally unique, previously unexplored hosts with plausible Ca-migration pathways for subsequent Ca$^{2+}$ mobility evaluation.

\subsection{Ca mobility in candidate frameworks}
\begin{figure}[h!]
    \centering
    \includegraphics[width=1\textwidth]{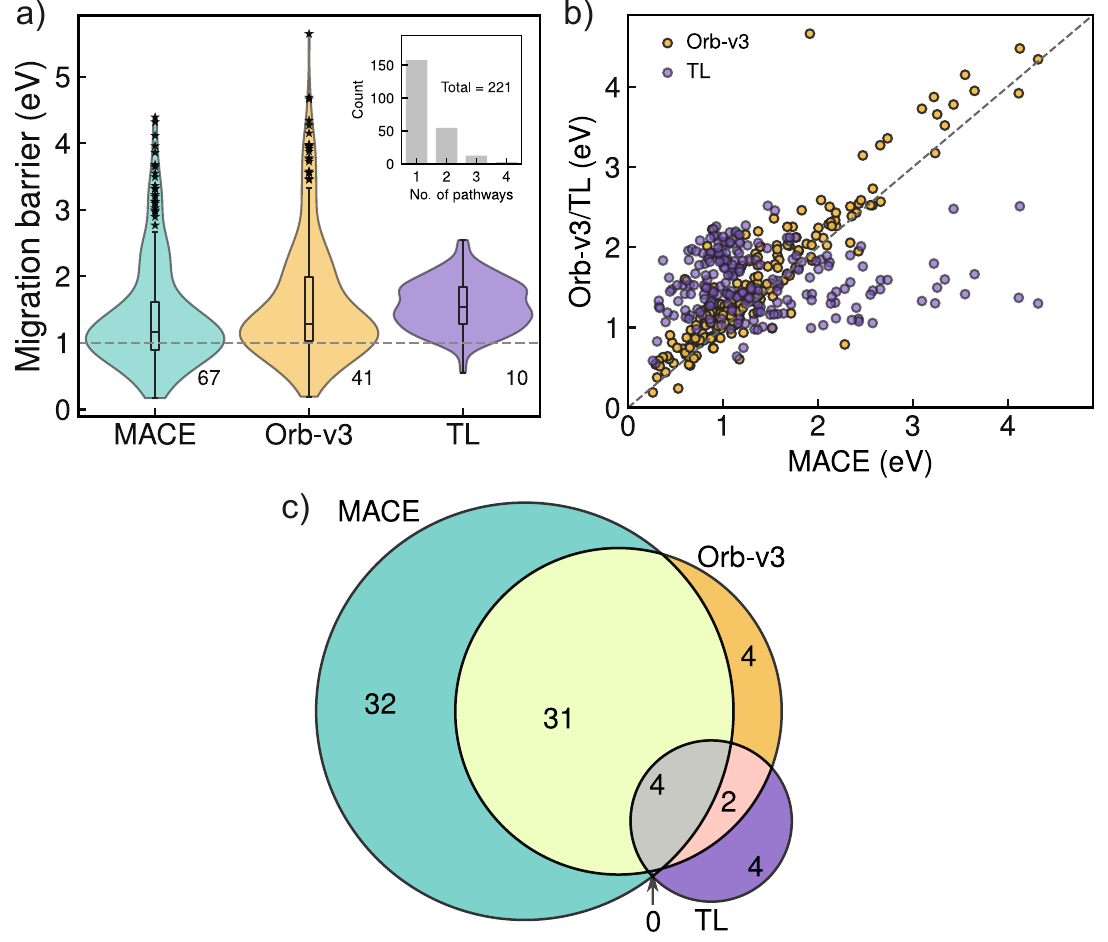} 
    \caption{
    Ca$^{2+}$ $E_m$ within screened frameworks, as predicted by MACE (teal), Orb-v3 (orange), and TL (purple). a) Violin plot showing the distribution of $E_m$; the inset bar chart reports the number of structures with one, two, three, and four distinct pathways. b) Parity plot comparing Orb-v3 and TL $E_m$ values against MACE. c) Venn diagram of the overlap in selected candidate frameworks among the three models.}
    \label{fig:barrier}
\end{figure}

\noindent Given the key role $E_m$ plays in estimating ionic mobility in materials, we show the distribution of Ca$^{2+}$ $E_m$ values across the screened and curated candidates in \textbf{Figure~\ref{fig:barrier}a}. $E_m$ predictions by MACE, Orb-v3 and the TL model are indicated by  teal, orange, and purple violins, respectively. The inset bar plot in \textbf{Figure~\ref{fig:barrier}a} shows the number of compounds with one, two, three, or four distinct migration pathways. In total, the dataset comprises 221 candidate frameworks spanning 299 unique migration pathways. For frameworks with multiple migration pathways, the active transport network may involve one, some, or all of these paths. However, we represent each framework by the median $E_m$ across its distinct pathways for the purpose of considering statistical distribution of $E_m$. The embedded inner boxes within each violin spans the interquartile range (from the lower to the upper quartile), with the horizontal black line marking the median of the $E_m$ distribution and black stars denoting outliers. The dashed horizontal gray line represents the $E_m$ threshold of $\sim$1~eV that we have used to identify Ca-cathode candidates with reasonable rate performance, a liberal threshold that is consistent with previous studies.\cite{lu2021searching, tekliye2022exploration, tekliye2024fluoride} The number of candidates predicted to lie within this $E_m$ threshold is indicated as text corresponding to each model.

With MACE predictions, the Ca$^{2+}$ $E_m$ span from 0.168 to 4.393~eV, with a median of 1.170~eV, indicating that the distribution is centered near 1~eV. Below the $E_m \le 1$~eV threshold, MACE identifies 67 frameworks with reasonable Ca-mobility, representing the largest pool of promising Ca-cathode candidates among the three models considered. With respect to Orb-v3, its $E_m$ predictions span a wider range than MACE (0.191--5.647~eV) with a higher median (1.281~eV), resulting in a $E_m$ distribution that is broadly centered near $\sim$1~eV but systematically shifted toward higher values with a pronounced tail beyond 3.5~eV. Under the $E_m \le 1$~eV threshold, Orb-v3 selects 41 candidates, which is lower than the 67 identified by MACE. Finally, our TL model predicts a $E_m$ range of 0.553--2.544~eV that is narrower than MACE and Orb-v3 but with a higher median of 1.535~eV, yielding only 10 candidates below the 1~eV threshold. The distribution of $E_m$ values exhibited by the TL model is also qualitatively different from MACE and Orb-v3, as highlighted by the difference in the shape of the purple violin compared to the teal and orange violins in \textbf{Figure~{\ref{fig:barrier}}a}. Overall, the three models differ markedly in selectivity: MACE is the most permissive, retaining the largest number of frameworks under the 1~eV threshold, whereas the TL model is the most conservative.

\textbf{Figure~\ref{fig:barrier}b} compares the Ca$^{2+}$ $E_m$ predicted by Orb-v3 (orange circles) and the TL model (purple) against those obtained from MACE. A significant number of points cluster near the parity line (dashed grey), indicating that all three models capture moderately similar trends in $E_m$ across the candidate set. Notably, Orb-v3 exhibits an almost linear correlation with MACE over the full $E_m$ range, with a marginal tendency to predict higher values, consistent with our observations in \textbf{Figure~{\ref{fig:barrier}}a}. In contrast, the TL model offers a more conservative $E_m$ landscape, assigning higher values at low–to-intermediate MACE-$E_m$ and lower values at high MACE-$E_m$. Thus, the TL model does exhibit qualitative differences compared to both MACE and Orb-v3, and identifies fewer candidates with Ca-mobility, which is consistent with its higher precision in identifying `good' conductors.\cite{devi2026leveraging} Nevertheless, the TL model adds complementary information to MACE/Orb-v3 predictions and supports a more robust estimation of Ca$^{2+}$ mobility across the candidate space.
    
To avoid over-reliance on any single model among MACE, Orb-v3, and TL for $E_m$ predictions, we use an ensemble (or MoE) approach in which frameworks predicted favorably by at least two of the three models are considered as candidate Ca-cathodes, with a subset of them validated with DFT-NEB subsequently. \textbf{Figure~\ref{fig:barrier}c} illustrates the overlap among the model-selected candidates. Applying the $E_m$ threshold yields 67, 41, and 10 candidates for MACE, Orb-v3, and TL, respectively. Since each model selects distinct, non-overlapping frameworks and the pairwise overlaps among the models differ, no single model is an obvious arbiter. Notably, all three models share four common candidates (gray region). The overlap between MACE and Orb-v3 comprises 35 candidates in total (31 shared only by MACE and Orb-v3 and the four shared by all models; light yellow and gray regions). MACE and TL share four candidates, all of which are in the three-way overlap. Orb-v3 and TL share six candidates, i.e., two shared only by these two models plus the four shared by all three.  Therefore, applying the ensemble approach, we arrive at a final set of 37 candidate Ca-cathode frameworks (green, gray, and pink regions in \textbf{Figure~\ref{fig:barrier}}c) that should exhibit reasonable Ca-mobility.

\subsection{DFT-NEB validation}
\begin{figure}[h!]
    \centering
    \includegraphics[width=1\textwidth]{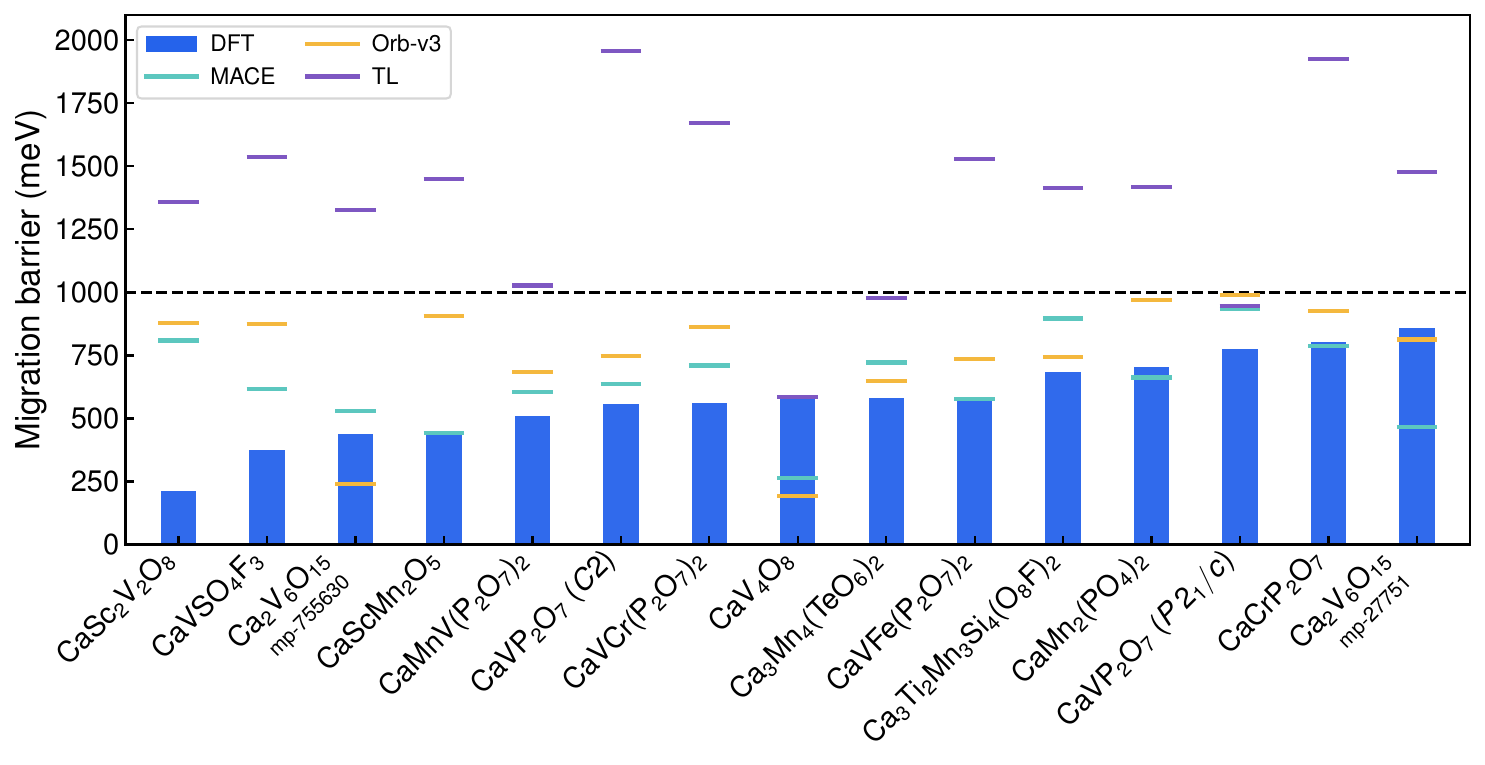}
    \caption{DFT–NEB $E_m$ for a subset of final candidate frameworks, shown alongside predictions from MACE (teal), Orb-v3 (orange), and TL (purple) overlaid as horizontal lines on each bar. The dashed black line marks the 1000 meV threshold. Space groups of select structures are indicated in parentheses. Compounds with the same chemical formula and space group but different StructureMatcher assignments are distinguished by their MP IDs.}
    \label{fig:dft-barrier}
\end{figure}

\noindent To reduce computational cost associated with DFT-NEB calculations and validate as many of the ensemble-model-predicted Ca-cathode candidates, we excluded structures sharing the same composition and space group that are classified as distinct by \texttt{pymatgen}'s \texttt{StructureMatcher} due to minor atomic rearrangements, except Ca$_2$V$_6$O$_{15}$. This reduction yields 26 distinct frameworks for DFT-NEB calculations. \textbf{Figure~\ref{fig:dft-barrier}} presents a bar plot of DFT-estimated $E_m$ alongside the corresponding MACE, Orb-v3, and TL predictions for a subset containing 15 frameworks (14 distinct frameworks + 1 additional Ca$_2$V$_6$O$_{15}$) for which DFT-NEB calculations converged, namely, 
    $\mathrm{CaSc_2V_2O_8}$,
    $\mathrm{CaVSO_4F_3}$, 
    $\mathrm{Ca_2V_6O_{12}}$ (MP ID: mp-755630)
    $\mathrm{CaScMn_2O_5}$,  
    $\mathrm{CaMnV(P_2O_7)_2}$, 
    $\mathrm{CaVP_2O_7}$ (\textit{C2}), 
    $\mathrm{CaVCr(P_2O_7)_2}$, 
    $\mathrm{CaV_4O_8}$,
    $\mathrm{Ca_3Mn_4(TeO_6)_2}$, 
    $\mathrm{CaVFe(P_2O_7)_2}$,  
    $\mathrm{Ca_3Ti_2Mn_3Si_4(O_8F)_2}$, 
    $\mathrm{CaMn_2(PO_4)_2}$,
    $\mathrm{CaVP_2O_7}$ ($\mathrm{P}\,2_1/\mathrm{c}$),
    $\mathrm{CaCrP_2O_7}$, and
    $\mathrm{Ca_2V_6O_{15}}$ (MP ID: mp-27751). 
Polymorphs with the same composition are distinguished by listing their space group from the original MP structure in parentheses. In the case of Ca$_2$V$_6$O$_{15}$, we include two structures classified as distinct by \texttt{StructureMatcher} despite identical composition and space group, for additional validation. The DFT-NEB-calculated minimum energy pathways of Ca$^{2+}$ migration for the 15 frameworks are compiled in \textbf{Figure~S2}. Note that our DFT-NEB calculations did not successfully converge for the remaining 12 frameworks.

Importantly, the DFT-computed $E_m$, for all frameworks whose calculations converged, lie within the tolerance threshold of $E_m < 1000$ meV (black dashed line in \textbf{Figure~\ref{fig:dft-barrier}}) adopted in this study, validating the ability of our model ensemble to reliably identify promising Ca cathode candidate frameworks. To gain further confidence in our ensemble predictions, we expanded the validation by randomly sampling a subset of three non-shortlisted candidates, and found good agreement with DFT-NEB calculations for such structures as well (see \textbf{Figures~S3} and \textbf{S4}). Among the ML models, MACE (teal lines in \textbf{Figure~\ref{fig:dft-barrier}}) shows the best agreement with DFT, yielding a mean absolute error (MAE) against all DFT-calculated (including the non-candidates) $E_m$ of 161~meV and closely reproducing the overall trend of the barriers across the dataset. Orb-v3 (orange lines) exhibits larger deviations with an MAE of 274~meV, generally capturing the qualitative trends but often overestimating $E_m$. Interestingly, the TL model (purple lines) shows substantially larger discrepancies, with an MAE of 671~meV, and a tendency to significantly overestimate $E_m$. Taken together, these results suggest that MACE reproduces the DFT $E_m$ more closely compared to the other models considered.

Notably, we identify CaSc$_2$V$_2$O$_8$ and CaVSO$_4$F$_3$ to exhibit the lowest DFT-calculated $E_m$ of 213 and 376~meV, respectively, with both values being two of the lowest Ca$^{2+}$ $E_m$ reported in the literature so far. Other candidate structures exhibiting DFT-NEB $E_m <$500~meV, which is an indicator of facile Ca$^{2+}$ motion under typical electrochemical conditions, are CaScMn$_2$O$_5$ (447~meV), and Ca$_2$V$_6$O$_{15}$ (MP ID: mp-755630; 438~meV). Given the systematic (and qualitative) agreement between the ensemble-classification of structures with reasonable Ca-mobility and low $E_m$ with DFT, we consider the full set of 37 frameworks identified by our MoE predictions to be promising Ca-cathode candidates. The 37 frameworks do contain multiple polymorphs of the same composition, thus resulting in a total of 25 distinct and promising Ca-cathode compositions. We summarize the Ca$^{2+}$ $E_m$ calculated using MACE, Orb-v3, TL, and/or DFT for the 37 identified Ca cathode frameworks, along with their compositions, space groups, MP-IDs, and other screening metrics computed in this work in \textbf{Table~\ref{tab:candidate-frameworks}}. Schematics of all 37 structures are compiled in \textbf{Figures~S5-S8} of the SI. Among the 25 promising compositions, we highlight Ca$_3$(CoO$_2$)$_4$, Ca$_3$Mn$_4$(TeO$_6$)$_2$, CaVF$_5$, and CaVSO$_4$F$_3$ due to their thermodynamic stability in the charged state ($E^\mathrm{hull} = 0$~meV/atom), signifying their thermodynamic accessibility and potentially successful experimental synthesis. 

\begin{table} 
\footnotesize
\centering
\caption{Summary of Ca-cathode candidates identified in this work. For each framework, we report the composition, space group, MP-ID, substituted site and the VPV, $E^{\mathrm{hull}}$ at discharged and charged compositions, average voltage versus Ca metal, and Ca$^{2+}$ $E_m$ obtained from DFT, MACE, Orb-v3, and TL. Two $E_m$ for a given framework indicate two distinct migration pathways. Missing DFT values correspond to polymorphs excluded or non-convergence of NEB calculations.}
\setlength{\tabcolsep}{0.5pt}  
\begin{tabular}{llccccccccr}
\hline
\hline
\multirow{2}{*}{Composition} & \multirow{2}{*}{Space group} & \multirow{2}{*}{MP-ID} & Substituted site & \multicolumn{2}{c}{$E^\mathrm{hull}$ (meV/atom)} & Average voltage & \multicolumn{4}{c}{Ca$^{2+}~E_m$ (meV)} \\
& & & (VPV; \AA$^3$) & Discharged & Charged & (V vs. Ca) & DFT & MACE & Orb-v3 & TL \\
\hline
Ca$_3$(CoO$_2$)$_4$ & $Pmmn$ & mp-766225 & Na  (13.29) & 99 & 0 & 4.5 &  & 307 & 377&827 \\
Ca$_3$Mn$_4$(TeO$_6$)$_2$ & $Pnma$ & mp-561325 & Na  (14.07) & 71 & 0 & 4.1 & 594 & 722 & 647&978 \\
CaCo$_3$NiO$_8$ & $R3m$ & mp-2219095 & Mg  (13.88) & 68 & 43 & 3.4 &  & 292 & 538&553 \\
\multirow{6}{*}{CaMnV(P$_2$O$_7$)$_2$} & \multirow{6}{*}{$P1$} & mp-764481 & Li (13.26) & 38 & \multirow{6}{*}{16} & \multirow{6}{*}{3.8} &  & 640 & 638& 1110 \\
& & mp-764622 & Li  (13.33) & 36 & & & & 759 & 841 & 1190 \\
& & mp-764488 & Li  (13.34) & 36 & & & 511 & 604 & 685 & 1215\\
& & mp-764484 & Li  (13.36) & 39 & & & & 982 & 985 & 1027\\
& & mp-759232 & Li  (13.51) & 36 & & & & 733 & 833 & 1228\\
& & mp-764480 & Li  (13.36) & 38 & & & & 872 & 862 & 1061\\
\multirow{2}{*}{CaVP$_2$O$_7$} & $P\,2_1/c$ & mp-758360 & Li  (14.28) & 12 & 10 & 2.9 & 776 & 934 & 988 & 1933\\
& $C2$ & mp-2217313 & Mg  (13.47) & 6  & 36 & 3.1 & 558 & 636 & 749 & 1958\\
\multirow{4}{*}{CaVFe(P$_2$O$_7$)$_2$}  & \multirow{4}{*}{$P1$} & mp-752490 & Li  (13.55) & 21 & 24 & 3.8 & 583 & 576 & 736 & 1530\\
& & mp-1235847 & Li  (14.53) & 24 & 24 & 3.7 &  & 801 & 842 & 1526 \\
& & mp-765981 & Li  (13.32) & 24 & 24 & 3.7 &  & 652 & 750 & 1535\\
& & mp-1235226 & Li  (13.20) & 30 & 25 & 3.7 &  & 433 & 608 & 1622\\
CaCoP$_2$O$_7$ & $P1$ & mp-1569230 & Li  (14.11) & 46 & 59 & 4.4 &  & 973 & 869 & 1514 \\
CaCrP$_2$O$_7$ & $P\,2_1$ & mp-19103 & Li  (13.31) & 34 & 41 & 3.4 & 802 & 787 & 925 & 1927 \\
CaVPO$_5$ & $Pnma$ & mp-760911 & Na  (13.40) & 84 & 29 & 3.1 &  & 416 & 641 & 1186\\
\multirow{3}{*}{CaVF$_5$} & \multirow{3}{*}{$P\,2_1/c$} & \multirow{2}{*}{mp-759186} & \multirow{2}{*}{Li  (13.24)} & \multirow{2}{*}{81} & \multirow{2}{*}{0} & \multirow{2}{*}{4.0} &  & 827 & 753 & 1777 \\
& &  & & &  &  & & 948 & 987 & 2080\\
& & mp-759086 & Li  (13.38) & 88  & 24 & 4.2 &  & 642 & 539 & 1843  \\
CaVSO$_4$F$_3$ & $P1$ & mp-769603 & Li  (14.02) & 11 & 0 & 4.1 & 376 & 618 & 875 & 1586 \\
CaVCr(P$_2$O$_7$)$_2$ & $P1$ & mp-766000  & Li  (13.41) & 37 & 41 & 3.6 & 560 & 710 & 864 & 1673\\
CaV$_3$Fe$_2$CuO$_{12}$ & $P1$ & mp-766078 & Li  (14.50) & 42 & 60 & 3.6 &  & 901 & 740 & 1441\\
Ca$_3$Ti$_2$Mn$_3$Si$_4$(O$_8$F)$_2$ & $P\,2_1/c$ & mp-1210891 & Na  (13.86) & 33 & 81 & 2.9 & 686 & 896 & 744 & 1415\\
Ca$_2$V$_4$OF$_{12}$ & $P1$ & mp-757416 & Li  (13.55) & 60 & 59 & 3.1 &  & 674 & 856 & 1978\\
CaScMn$_2$O$_5$ & $Cmcm$ & mp-770412 & X  (13.59) & 90  & 58 & 2.5 & 447 & 443 & 905 & 1450\\
Ca$_2$V$_9$O$_{22}$ & $P\,2_1/m$ & mp-849581 & X  (13.50) & 68 & 52 & 3.4 &  & 335 & 517 & 1244\\
\multirow{2}{*}{Ca$_2$Mn$_5$(FeO$_6$)$_2$} & \multirow{2}{*}{$Cc$} & \multirow{2}{*}{mp-769550} & \multirow{2}{*}{Li (13.47)} & \multirow{2}{*}{71} & \multirow{2}{*}{99} & \multirow{2}{*}{3.2} &  & 321 & 335 & 1059\\
& & &  & & &  &  & 472 & 550 & 1077\\
CaMnO$_2$ & $P6\,_3/mmc$ & mp-971647 & Na  (13.75) & 84 & 78 & 2.5 &  & 1216& 936&747 \\
CaMnCSO$_7$ & $P2\,_1$ & mp-754499 & Li  (13.43) & 44 & 7 & 4.0 &  & 1519 & 990&997 \\
\multirow{2}{*}{Ca$_2$V$_6$O$_{15}$}  & \multirow{2}{*}{$C2/m$} &	mp-27751 	& Li 	(13.38) &	65 & 	57 & 	\multirow{2}{*}{3.4} & 859 &	529 &	240 &	1328 \\ 
 &	 & 	mp-755630 &	Li 	(14.48) &	55 & 	56 &  &	438 & 465 &	813 &	1476 \\
CaFe$_4$(PO$_4$)$_4$ &	$P1$ &	mp-1234013 &Mg 	(13.61) & 	42 & 	41 & 	3.2 &  & 482 &	554 &	1150 \\
CaMn$_2$(PO$_4$)$_2$ &	$P\bar{1}$ &	mp-1176722 &	Li 	(13.80) &	48 & 	34 & 	3.5 & 704 & 662 & 	969 &	1417 \\
Ca$_5$V$_{18}$O$_{44}$ &	$Pm$ &	mp-773027 &	Li 	(13.98) &	54 & 	52 & 	3.4 & 	& 325 & 589 & 	1334 \\
CaSc$_2$V$_2$O$_8$ & 	$C2$ &	mp-2218897 &	Mg 	(13.22) & 	76 & 	34 & 	2.6  &	213 & 809 & 	877 & 	1357 \\
CaV$_4$O$_8$ &	$C2/m$ &	mp-2220145 &	Mg 	(13.89) &	89 & 	64 & 	2.5 & 576	& 262 &	191 &  585 \\
\hline
\hline
\end{tabular}
\label{tab:candidate-frameworks}
\end{table} 

\section{Discussion}
\begin{figure}[h!]
    \centering
    \includegraphics[width=0.6\textwidth]{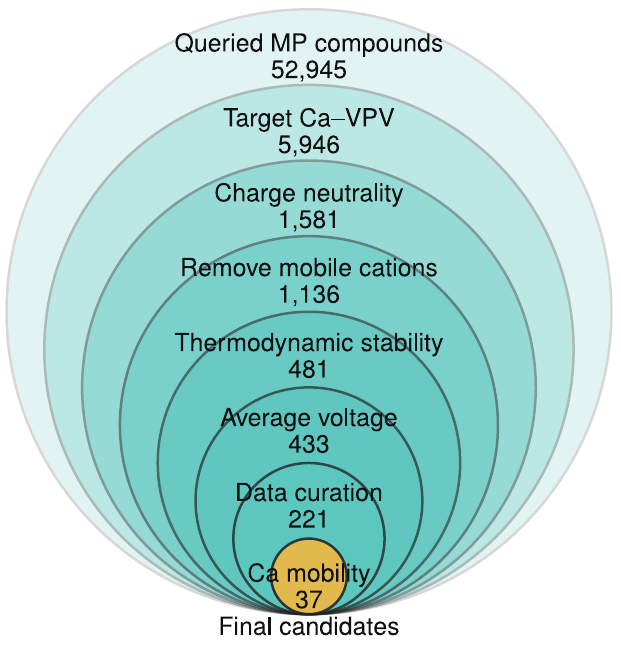}
    \caption{Summary of the Ca-cathode screening process. The concentric circles show the sequential filtering of structures and the number of compounds remaining after each screening step.}
    \label{fig:screening}
\end{figure}

\noindent In this work, we have performed a large-scale screening of potential Ca battery cathodes across 52,945 non-Ca-containing MP structures using a high-throughput workflow (\textbf{Figure~{\ref{fig:workflow}}}) accelerated by foundational ML models. We have applied a series of chemically and physically motivated filters, namely, geometry-based VPV (\textbf{Figure~{\ref{fig:mp-distribution}}}), electrostatic charge neutrality and presence of non-Ca mobile cations (\textbf{Figure~{\ref{fig:charge_remove}}}), thermodynamic (meta)stability and average intercalation voltage (\textbf{Figure~{\ref{fig:therm_stability}}}), duplicate structures (\textbf{Figure~{\ref{fig:deduplicate}}}), and Ca$^{2+}$ mobility (\textbf{Figure~{\ref{fig:barrier}}}), to down-select a final list of 37 distinct structures spread across 25 distinct compositions. Importantly, a section of our Ca-mobility estimates have been validated with computationally-heavy DFT-NEB calculations (\textbf{Figure~{\ref{fig:dft-barrier}}}), which are in good agreement with our ensemble approach of selecting candidates with reasonable $E_m$. We illustrate the overall screening workflow in \textbf{Figure~\ref{fig:screening}}, with \textbf{Table~\ref{tab:candidate-frameworks}} and \textbf{Figures~S5-S8} compiling all the candidates identified.

In our screening strategy, the Ca–VPV descriptor was constrained by a relatively strict tolerance window of $\pm 5\%$ around the Ca–VPV median (\textbf{Figure~{\ref{fig:mp-distribution}}}). This strict choice prioritizes frameworks whose local environments are most similar to those in Ca-containing frameworks, while keeping the number of candidates manageable after applying all subsequent filters. In principle, the tolerance window can be broadened to include additional candidates. However, as one moves progressively farther from the Ca–VPV median, the likelihood of identifying candidate frameworks that are thermodynamically (meta)stable is expected to decrease, possibly reducing overall screening efficiency. Based on the final set of candidates listed in \textbf{Table~\ref{tab:candidate-frameworks}}, the substituted cations/sites are predominantly Li, Na, Mg, and X, in descending order, suggesting that these species frequently provide structurally compatible host environments for Ca substitution and may therefore serve as practical starting points for Ca-cathode framework design. We note, however, that the prominence of Li-, Na-, and Mg-containing frameworks partly reflects their higher representation within the MP database, which increases their likelihood of appearing in the screened candidate set.

We evaluated charge neutrality (\textbf{Figure~{\ref{fig:charge_remove}}}) and thermodynamic stability (\textbf{Figure~{\ref{fig:therm_stability}}}) only for the fully charged and fully discharged compositions for any candidate framework, as explicitly sampling intermediate compositions would require exhaustive enumeration of Ca-vacancy configurations, which is computationally prohibitive in a high-throughput setting. Although such intermediate-state sampling could reveal additional charge-neutral and (meta)stable phases and thereby expand the pool of promising candidates, we do not capture any partially (de)intercalated compositions here, highlighting the need for more efficient methods to approximate intermediate-composition energetics in any follow-up large-scale screenings of Ca cathodes and beyond. Note that our average voltage calculations are also reported across the entire Ca-composition range (i.e., fully charged to fully discharged), where local variations in the voltage with Ca (de)intercalation can result in possible bottlenecks in the practical use of any candidate with electrolytes. One possible strategy to account for partial Ca compositions is to focus on only thermodynamically stable (i.e., $E^\mathrm{hull} = 0$~meV/atom) charged or discharged compositions (so that the likelihood of experimental synthesis is high) and perform a sub-sampling of intermediate Ca compositions by partial (de)intercalation, to reduce the compositional complexity.
    
One of the key bottlenecks in our screening workflow is the structural analysis required for assessing Ca$^{2+}$ mobility, which presently involves visually inspecting each structure to identify all plausible migration pathways (\textbf{Figure~{\ref{fig:deduplicate}}}). This step quickly becomes intractable in a truly high-throughput setting, where hundreds to thousands of candidate frameworks may need to be analyzed, and therefore limits the overall speed and scalability of the workflow. Developing robust, automated tools to systematically generate and rank migration pathways would thus be highly valuable for the community, enabling more efficient screening of potential ionic conductors and facilitating the extension of similar workflows to even larger chemical and structural design spaces.

Since a rigorous validation of the $E^\mathrm{hull}$ for any screened Ca-cathode would require computing ground-state energies for all competing phases in the relevant chemical space (elemental, binary, ternary, quaternary, etc.), which can be computationally prohibitive in a high-throughput setting, we did not explicitly validate the thermodynamic stability of the identified candidates with DFT. However, most of the screened frameworks lie only marginally above the convex hull (typically $\le 50$~meV/atom, as shown in \textbf{Table~\ref{tab:candidate-frameworks}}), and state-of-the-art ML models have been extensively benchmarked for accurate prediction of formation energies in inorganic systems.\cite{deng2025systematic, merchant2023scaling, riebesell2025framework} Therefore, we expect that the majority of the screened candidates will remain thermodynamically accessible and, in principle, synthesizable. For the final set of candidates identified by the ensemble of models (\textbf{Figure~{\ref{fig:barrier}}}), we further validated the transport properties for a subset of 15 candidates by explicitly computing Ca $E_m$ using DFT-based NEB calculations (\textbf{Figure~{\ref{fig:dft-barrier}}}), thus increasing the confidence of our predictions. Given our limited dataset of 15 DFT-NEB calculated $E_m$ values, we observe MACE to provide the closest agreement to DFT-values compared to Orb-v3 and TL.

While geometric descriptors such as the VPV provide an efficient and scalable way to identify geometrically-compatible Ca insertion sites, other approaches may also offer complementary insights for Ca-cathode discovery. In particular, insertion-site prediction based on the electronic charge density of the empty host framework has been proposed recently for Li-ion cathodes, where favorable cationic positions are identified by locating regions of low electronic charge density within the host lattice.\cite{shen2020charge} Since cations tend to occupy electrostatic potential minima, this approach provides a chemically informed route to identify stable insertion sites without prior knowledge of the fully intercalated structure. Such charge-density–guided insertion strategies may be particularly relevant for multivalent ions, such as Ca$^{2+}$, whose stronger electrostatic interactions with the host lattice play an important role in determining site stability. Exploring these electronic-structure–based descriptors alongside geometric approaches, such as VPV, could therefore provide additional opportunities for guiding the future design of battery electrode materials.

\section{Conclusion}
\noindent Accelerating the design and discovery of cathode frameworks capable of reversibly intercalating Ca$^{2+}$ is essential for advancing multivalent CBs as practical alternatives to LIBs. Here, we used a combination of geometric, electrostatic, thermodynamic, and kinetic descriptors to screen across 52,945 non-Ca-containing structures within the MP database and identify 37 promising and distinct structures spanning across 25 different cathode compositions. Specifically, we introduced a geometry-based screening metric that evaluates the compatibility of potential Ca$^{2+}$ sites within a structure using local VPV derived from Voronoi tessellation, providing a purely geometric and therefore transferable metric that avoids the limitations of ionic-radius–based heuristics across diverse bonding environments. Building on the geometric-screening and excluding structures that are non-charge-neutral and contain non-Ca mobile cations, we develop a scalable high-throughput framework in which foundational MLIPs (MACE in this work) accelerates the evaluation of ground-state energetics for thermodynamic stability and average voltage calculations, thereby resulting in a swift filtering of 433 candidates. For the evaluation of Ca$^{2+}$ $E_m$, we used a combination of ML models, including MACE, Orb-v3, and TL, in a consensus ensemble approach to increase the confidence and robustness of $E_m$ predictions. Notably, we selected promising candidates as those identified by at least 2 models to have $E_m$ lower than our threshold of 1~eV, resulting in the filtering of 37 promising Ca-cathode frameworks. Finally, we corroborate our ensemble-selected Ca-cathodes with DFT-NEB calculations on a representative subset, thus validating the robustness of our selection and accuracy of $E_m$ predictions. Importantly, our screening identifies two cathode candidates with swift Ca$^{2+}$ mobility (CaSc$_2$V$_2$O$_8$, and CaVSO$_4$F$_3$) and four candidates with thermodynamic stability in their charged state (Ca$_3$(CoO$_2$)$_4$, Ca$_3$Mn$_4$(TeO$_6$)$_2$, CaVF$_5$, and CaVSO$_4$F$_3$). Overall, our study demonstrates an effective strategy for accelerating Ca-cathode discovery and highlights the value of integrating transferable geometric descriptors, foundational ML models, and ensemble decision-making for materials exploration across broad chemical and structural spaces, paving the way for accelerated discovery in other intercalation chemistries and applications beyond batteries.

\section*{Acknowledgments}
\noindent G.S.G. acknowledges the financial support received from the Science and Engineering Research Board (SERB), Department of Science and Technology, Government of India, under sanction number IPA/2021/000007. D.B.T. and A.K.B. acknowledge financial assistance from the Indian Institute of Science (IISc) and the Ministry of Education, Government of India, respectively. The authors acknowledge the computational resources provided by the Supercomputer Education and Research Centre (SERC) at IISc, which were instrumental in carrying out the first-principles calculations and high-throughput screening workflows presented in this thesis. The authors also gratefully acknowledge the J\"{u}lich Supercomputing Centre at Forschungszentrum J\"{u}lich, Germany, for access to the JURECA supercomputer under projects hpc-prf-emdft and hpc-prf-desal, which enabled some of the NEB calculations reported in this work.

\section*{Data and code availability}
\noindent All calculated data and the scripts used in this work are available freely online on our \href{https://github.com/sai-mat-group/high-throughput-ca-cathodes}{GitHub} repository.

\section*{Conflicts of interest}
\noindent The authors have no conflicts of interest to declare.


\newpage
\bibliographystyle{unsrt}
\bibliography{bibfile}

\end{document}